\documentclass[aps,prl,twocolumn,superscriptaddress]{revtex4-2}
\usepackage{amsmath,amssymb,graphicx,bbold}
\usepackage{graphicx}  
\usepackage{dcolumn}   
\usepackage{bm}        
\usepackage{color}
\usepackage{xcolor}
\usepackage{physics}
\usepackage{ulem}
\usepackage{subfigure}
\begin{document}

\title{Machine learning recognition of light orbital-angular-momentum superpositions}
\author{B. Pinheiro da Silva}
\email{braianps@gmail.com}
\affiliation{Instituto de F\'{i}sica, Universidade Federal Fluminense, 24210-346 Niter\'{o}i, RJ, Brazil}

\author{B. A. D. Marques}
\email{brunodortamarques@gmail.com}
\affiliation{Universidade Federal Rural do Rio de Janeiro, 26285-060 Nova Iguaçu, RJ, Brazil}

\author{R. B. Rodrigues}
\email{rafaelbellasrodrigues@gmail.com}
\affiliation{Instituto de F\'{i}sica, Universidade Federal Fluminense, 24210-346 Niter\'{o}i, RJ, Brazil}

\author{P. H. Souto Ribeiro}
\email{p.h.s.ribeiro@ufsc.br}
\affiliation{Departamento de F\'\i sica, Universidade Federal de Santa Catarina, 88040-900 Florian\'opolis, SC, Brazil}

\author{A. Z. Khoury}
\email{azkhoury@id.uff.br}
\affiliation{Instituto de F\'{i}sica, Universidade Federal Fluminense, 24210-346 Niter\'{o}i, RJ, Brazil}
\date{\today}

\begin{abstract}
We develop a method to characterize arbitrary superpositions of light orbital angular momentum (OAM) with high fidelity by using astigmatic transformation and machine learning processing. In order to define each superposition unequivocally, we combine two intensity measurements. The first one is the direct image of the input beam, which cannot distinguish between opposite OAM components. The second one is an image obtained using an astigmatic transformation that removes this ambiguity.  Samples of these image pairs are used to train a convolution neural network and achieve high fidelity recognition of arbitrary OAM superpositions with dimension up to five.
\end{abstract}

\maketitle

It is well known that orbital angular momentum of light (OAM) \cite{padgett} has many applications in different areas such as optical manipulation \cite{padgett2, gece}, communications \cite{cadu, ambrosio, tamburini, eutp, cadu10} and simulation \cite{Araujo18,Ribeiro20,Haffner20}. Along with this broad range of applications, there is the requirement of being able to prepare and identify OAM single and superposition modes.
To this end, interesting schemes can be realized with machine learning algorithms that have been employed to improve the state-of-the-art in computer vision and pattern recognition \cite{lecun2015nature}. In particular, some applications concern structured light, including recognition and correction of OAM beams distorted by propagation through a turbulent medium \cite{tub00, tub0, tub1, tub22, tub2, tub3, tub4, bhusal, tub5, wang, huang, tub8}, direct recognition of OAM modes \cite{safura, ff} and classification of vector vortex beams \cite{vvb}. However, these previous investigations could not resolve OAM superpositions including positive and negative topological charges  \cite{Mumtaz}, which is crucial for quantum information applications.

Pure OAM beams are characterized by rotationally symmetric intensity distributions and an azimuthal variation of the phase given by a topological charge $\ell\in \mathbb{Z}\,$. Opposite OAM values cannot be distinguished by a single direct intensity measurement and one must resort to inteferometric techniques \cite{maru, si} or astigmatic transformations \cite{holandeses,vaity} in order to distinguish between left- and right-handed beams. Moreover, determining the coefficients of arbitrary OAM superpositions, their weights and relative phases, is a difficult task. In fact, astigmatic transformations can be used to perform tomography of OAM qubits \cite{eu1}. However, three distinct intensity measurements were required and the method was limited to OAM spaces of dimension two. 

In this work, we combine astigmatic transformation and machine learning techniques to achieve high fidelity in the characterization of arbitrary superpositions in OAM spaces with dimensions up to five. Our method is based on two distinct intensity measurements: \textit{i}) direct image of the input beam, \textit{ii}) image of the converted beam after astigmatic transformation. We use a convolution neural network (CNN) to recover the weights and relative phases of the coefficients in the OAM superposition. The database used for training the CNN consists of theoretical and experimental images.

The transverse structure of paraxial beams propagating in free space can be described by Laguerre-Gaussian (LG) functions. For a beam with wave-number $k\,$, propagating along the $z$ axis, the LG function reads 
\begin{align}
&\textrm{LG}_{\ell,p}(r,\theta) = \frac{\mathcal{N}_{\ell p}}{w}\,
\tilde{r}^{\vert\ell\vert} \,\textrm{L}_p^{\vert\ell\vert}\left(\tilde{r}\right) \,e^{-\frac{\tilde{r}^2}{2}} 
e^{i\ell\theta}\,e^{-i\Phi_N}\,,
\\
&\Phi_N = \frac{k\,r^2}{2R} + (N+1) \arctan\left( z/z_0\right)\,,\qquad \tilde{r} = \sqrt{2}\, r/w\,,
\nonumber
\end{align}
where $N=2p+\vert\ell\vert$ is the mode order, $\ell$ is the topological charge, $p$ the radial number, $\textrm{L}_p^{\vert{\ell}\vert}$ are generalized Laguerre polynomials and $\mathcal{N}_{\ell p}$ is a normalization 
constant. The beam parameters are the wave-front radius $R$, the width $w$ and the Rayleigh length $z_0\,$, which also characterize the Hermite-Gaussian modes
\begin{align}
&\textrm{HG}_{m,n}(x,y) = \frac{\mathcal{N}_{mn}}{w}\,
\textrm{H}_m\left(\tilde{x}\right) \,\textrm{H}_n\left(\tilde{y}\right) \,e^{-\frac{\tilde{x}^2 + \tilde{y}^2}{2}} 
\,e^{-i\Phi_N}\,,
\\
&\tilde{x} = \sqrt{2}\, x/w\,,\qquad \tilde{y} = \sqrt{2}\, y/w\,,
\nonumber
\end{align}
where $\mathcal{N}_{mn}$ is the proper normalization constant and the HG mode order is $N=m+n\,$. 

Both the LG and HG modes constitute orthonormal and complete bases of the transverse mode vector space. This space can be cast as a direct sum of subspaces related to the different mode orders. For a given order N, it is possible to define a subspace of dimension D = N + 1 analogous to the Hilbert space of a qudit, which is a quantum D-level system. LG and HG modes up to order N are vectors in this space and are connected to each other by unitary transformations. In this sense, an arbitrary transverse mode qudit can be written as a superposition of LG modes of order $N$ according to
\begin{equation}\label{sup1}
\vert\psi\rangle_D= \sum_{\ell,p} 
c_{\ell,p}\,\vert\textrm{LG}_{\ell,p}\rangle \,,
\end{equation}
where the summation runs over indices $\ell$ and $p$ restricted by $2p+\vert\ell\vert=D-1$ and $c_{\ell,p} $ is a complex weight. 

\begin{figure}[h!]
	\includegraphics[scale=0.7]{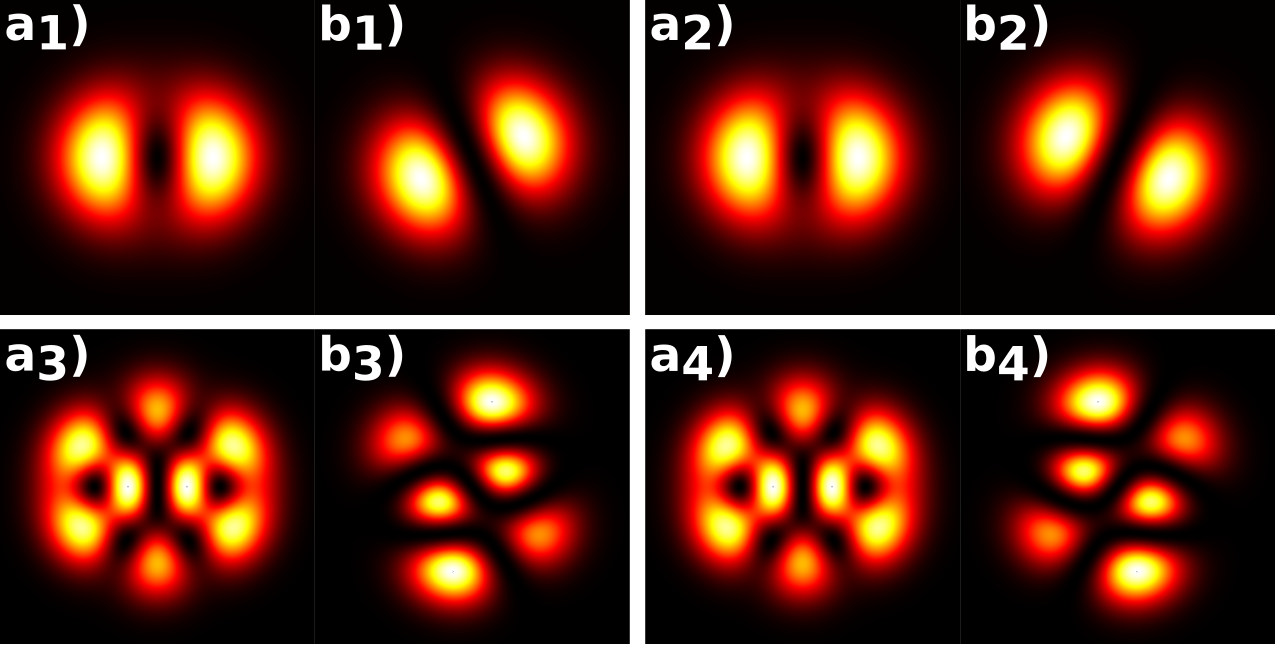}
	\caption{Theoretical intensity images of four different superpositions ($\vert\psi_1\rangle,\vert\psi_2\rangle,\vert\psi_3\rangle,\vert\psi_4\rangle$): a) are the direct images of the superpositions. b) are the images after the tilted lens. }
	\label{fig:mc}
\end{figure}

In view of the potential use of these superpositions in applications, their recognition is an essential task that can be approached in many ways, starting from the analysis of their intensity patterns. However, pure LG modes with the same values of $p$ and $\vert\ell\vert$ have identical intensity distributions, so it is impossible to distinguish them from a direct intensity measurement only. In fact, this problem is more general because any
superposition of the type defined by Eq.\eqref{sup1} is subjected to the following symmetry condition
\begin{eqnarray}\label{degeneracy}
\bigg\vert \sum_{\ell,p} 
c_{\ell,p}\,\textrm{LG}_{\ell,p}(\mathbf{r})\bigg\vert^2 
= \bigg\vert\sum_{\ell,p} 
c_{-\ell,p}\,\textrm{LG}_{\ell,p}(\mathbf{r})\bigg\vert^2\,.
\end{eqnarray}
This degeneracy can be lifted by supplementing the direct measurement with a second image obtained from astigmatic mode conversion of the input beam \cite{holandeses,eu1}. The mode converter acts as a unitary transformation restricted to each mode order subspace. Therefore, it can be written as the direct sum of $SU(D)$ operators:
\begin{eqnarray}\label{sup2}
MC &=& \sum_D^{\bigoplus} MC_D\;,
\\
MC_D &=& \sum_{m=0}^{D-1}e^{i(m-n)\frac{\pi}{4}}\,\vert\textrm{HG}_{m,n}\rangle\langle\textrm{HG}_{m,n}\vert 
\,,
\nonumber
\end{eqnarray}
with $n=D-m-1\,$. In Eq.\eqref{sup2} we made use of the Hermite-Gaussian base vectors $\{\vert\textrm{HG}_{m,n}\rangle\}$ 
which are the eigenmodes of the astigmatic transformation.
In Fig.\ref{fig:mc} the astigmatic method is illustrated for 4 different degenerate patterns.  
Images a\textsubscript{1}) and a\textsubscript{2}) display two degenerate intensity plots 
associated with the following superpositions
\begin{eqnarray}\label{psi12}
\vert\psi_1\rangle &=&\, 0.92\,\vert\textrm{LG}_{+1,0}\rangle + 0.38\,\vert\textrm{LG}_{-1,0}\rangle
\,,
\nonumber\\
\vert\psi_2\rangle &=&\,  0.38\,\vert\textrm{LG}_{+1,0}\rangle + 0.92\,\vert\textrm{LG}_{-1,0}\rangle
\,.
\end{eqnarray}
The degeneracy is lifted by the mode converted images b\textsubscript{1}) and b\textsubscript{2}).
We also illustrate the method with more complex patterns, such as those exhibited in images 
a\textsubscript{3}) and a\textsubscript{4}), corresponding to
\begin{eqnarray}\label{psi13}
\vert\psi_3\rangle  \!&=&\!
 \frac{1}{2}\left(\vert\textrm{LG}_{+3,0}\rangle - \vert\textrm{LG}_{-3,0}\rangle + \vert\textrm{LG}_{+1,1}\rangle
+ \vert\textrm{LG}_{-1,1}\rangle\right),
\nonumber\\
\vert\psi_4\rangle \!&=&\!
 \frac{1}{2}\left(-\vert\textrm{LG}_{+3,0}\rangle + \vert\textrm{LG}_{-3,0}\rangle + \vert\textrm{LG}_{+1,1}\rangle
+ \vert\textrm{LG}_{-1,1}\rangle\right).
\nonumber\\
\end{eqnarray}
Although they exhibit the same direct image pattern, they can be resolved by the mode converted 
images shown in b\textsubscript{3}) and b\textsubscript{4}).
Therefore, the two images are sufficient to define the mode superposition unequivocally, since they are capable of resolving opposite OAM states. 

The brute-force method for identifying a given superposition of modes, would be the following: i) perform the measurement of the intensity patterns of direct and mode-converted image; ii) generate theoretical intensity patterns with tentative mode superpositions and compare with the measured patterns; iii) use some optimization procedure to obtain the superposition that best approaches the measurement. However, more efficient strategies are available and we will demonstrate the use of machine learning to improve the recognition method.

We test these ideas experimentally. The experimental setup is shown in Fig. 2. A Gaussian beam from a Nd:YAG laser ($\lambda = 1064\, nm$) is sent to a spatial light modulator (SLM) programmed to produce an arbitrary mode superposition within the subspace of order $N$. The beam splitter (BS) is used to split the incoming beam, transmitting half of the intensity to the spherical lens L\textsubscript{a} ($f = 1\, m$) and forming the direct image in the CCD (charge-coupled device) camera. The reflected beam having the other half of the intensity is sent to mirror ($M$) and then to the tilted lens $L_b$ ($f = 1\, m$) that performs the astigmatic transformation. The beams are acquired in a single frame by the camera positioned at a distance of $0.74\, m$ from both lenses. 
\begin{figure}[h!]
	\includegraphics[scale=0.25]{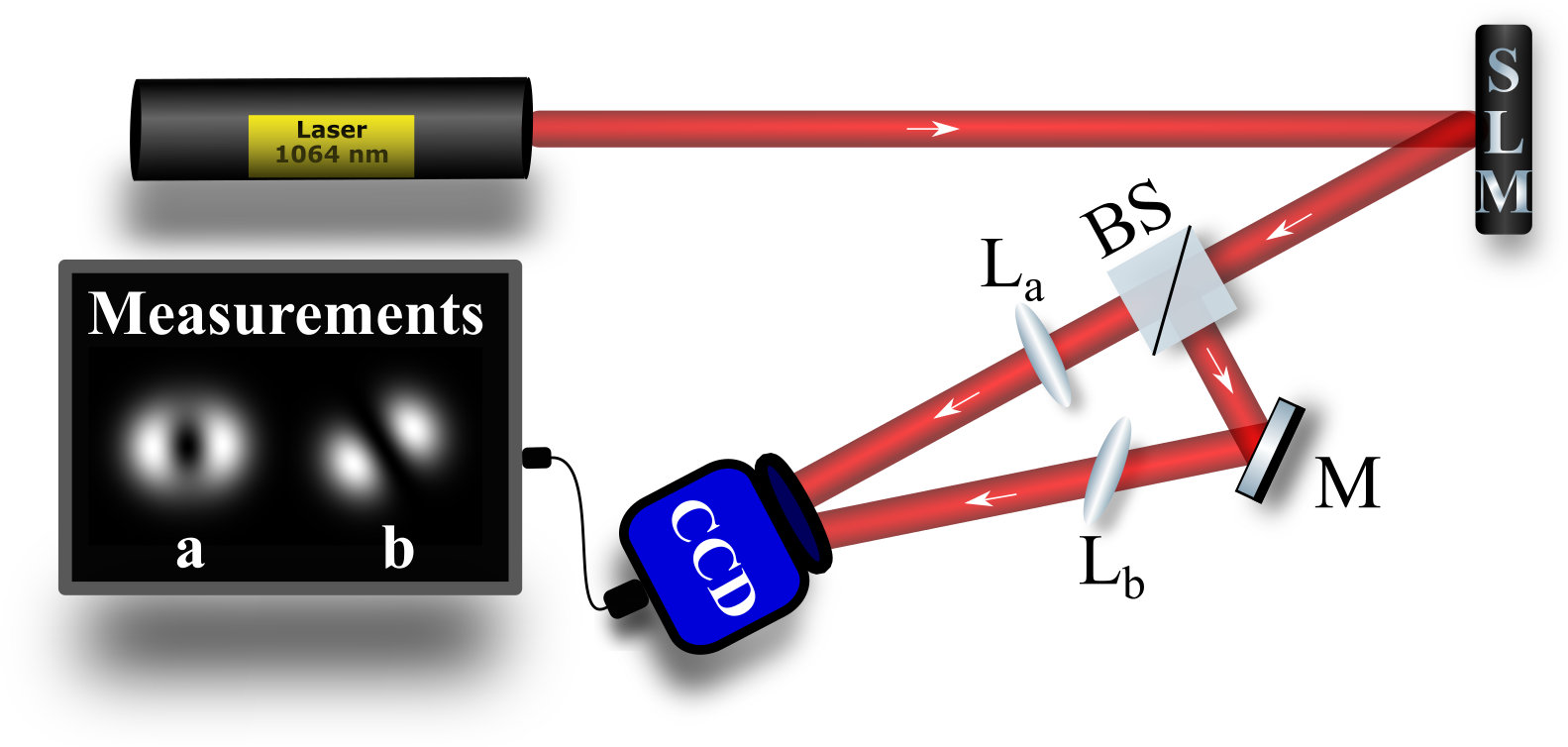}
	\caption{Experimental setup.}
	\label{fig:setup}
\end{figure}

To recover the complex weights $c_{\ell,p}$ of a given superposition, we employ a deep learning method denominated convolutional neural network (CNN), which is appropriated for image processing. Unlike traditional machine learning algorithms, the CNN can automatically select and extract key-features of images to solve pattern recognition tasks. This representation-learning ability \cite{bengio2013representation}, combined with the available processing power of modern graphics processing units (GPUs), allows the usage of a large number of images to construct a robust recognition system.%
\begin{figure}[h!]
	\includegraphics[scale=0.23]{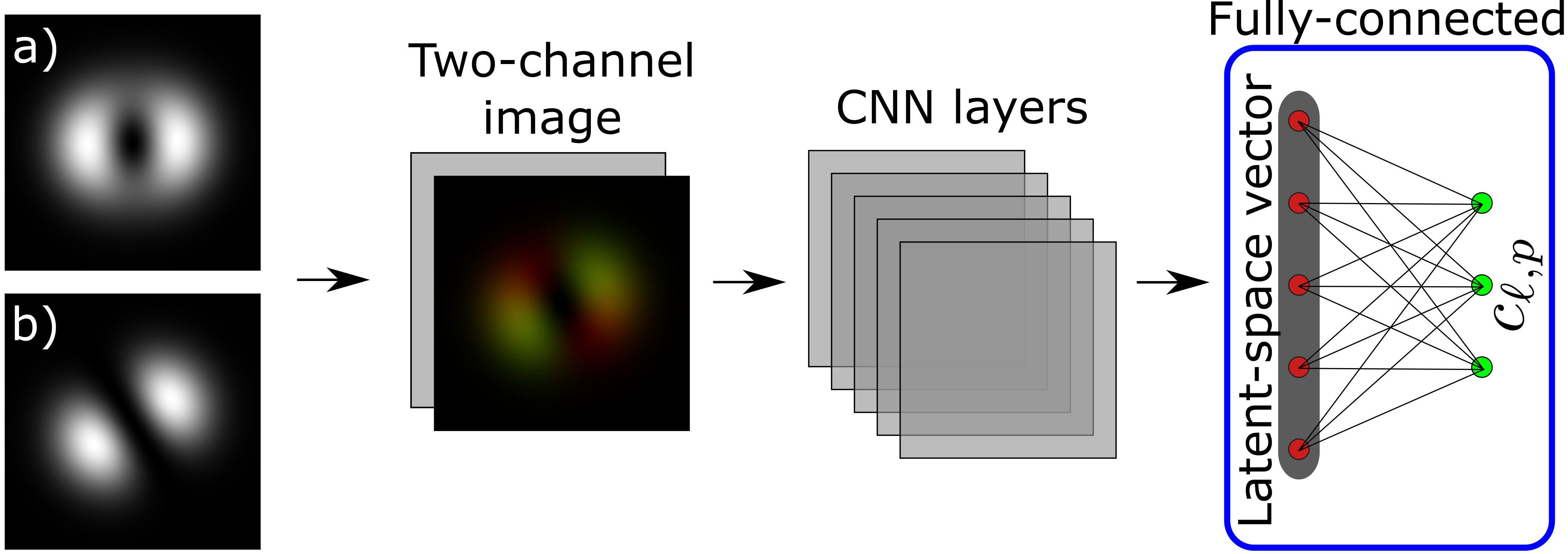}
	\caption{Superposition recognition system. The system receives two intensity images (a, b) and combines them in a two-channel image fed to the CNN. The CNN extracts a latent-space feature vector that is used to estimate the weights $c_{l,p}$ of the superposition.}
	\label{fig:cnn}
\end{figure}

The recognition system developed in this work takes the two input images of the mode superposition (direct and transformed), and outputs the coefficients $c_{\ell,p}$ of Eq. \eqref{sup1}.  We employ a 34-layer CNN that uses a series of convolutions and non-linear functions to extract the images' features. Then, these features are stored into a latent-space vector representing the most relevant features for recognition of the mode superposition. The latent-space vector is fed to the estimator (a fully-connected layer) that outputs a vector representing the values of the $c_{\ell,p}$.  Fig. \ref{fig:cnn} shows the overall representation of our system.

The architecture of our CNN is based on the residual neural network \cite{he2016deep}. We use 16 residual blocks totaling 32 convolution layers with ($3 \times 3$) kernel-size, an initial convolution layer with ($7 \times 7$) kernel-size, and a fully connected layer with 512 units. We state the recognition problem as a regression task, in which the CNN estimates a numerical value for the $c_{\ell,p}$. 

Typically, CNNs are trained using big datasets \cite{imagenet, coco}. In a recent work \cite{Mumtaz} superpositions of LG modes with $p=0$ and $ 0 \leq \ell \leq 9$ used $10^5$ theoretical samples, which is a relatively small dataset. However, all topological charges have the same sign, which facilitates the mode recognition. We choose to use the same order of magnitude for the datasets in our experiments. For each dimension $D\,$, we produce a dataset with $(D-1)\times10^4$ experimental samples of arbitrary superpositions. Moreover, we include both theoretical and experimental samples. The dataset is split into three parts: $75\,\%$ for the training, $15\,\%$ for the validation, and $10\,\%$ for the test. 

In Machine learning, the CNN model is a function composed of convolution operations that tries to fit the data based on previous examples during the training process. The CNN training process consists of providing examples from the training dataset and adjusting the CNN's parameters regarding a loss function. A training epoch is defined by the processing of all the examples in the training dataset. We validate the training epoch by evaluating the examples from the validation dataset. To ensure the CNN model's generality, we test the trained model with the testing dataset, which consists of images never used before during previous training and validation steps.

Our CNN initializes with randomly sampled parameters. The network is trained for $100$ epochs or until convergence using the Adam optimizer \cite{AdamKingma14} with a learning rate of $0.001$. We consider that the model converges if the validation loss does not improve after $40$ epochs. The loss function employed for training the CNN is $1- \mathcal{F}$, where $\mathcal{F}=\vert  \langle \psi_{D_G} \vert\psi_{D_E}\rangle\vert^2$ is the fidelity between the ground truth state $\vert\psi_{D_G}\rangle$ and the estimated state $\vert\psi_{D_E}\rangle\,$.

\begin{figure}[h!]
	\includegraphics[scale=0.55]{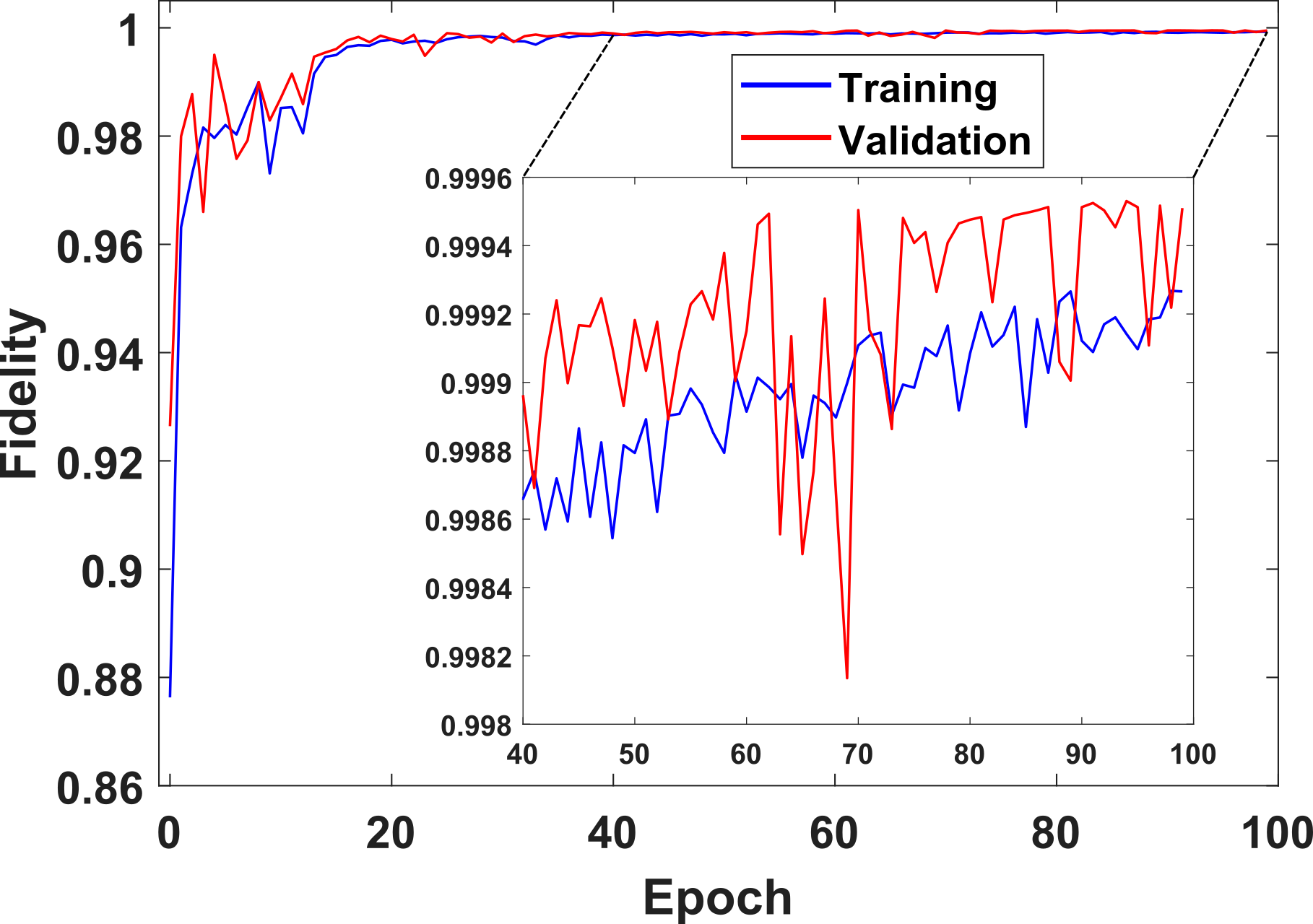}
	\caption{Mean fidelity in each epoch for training and validation process in the case  $D=2$.}
	\label{fig:d1}
\end{figure}

Figure \ref{fig:d1} shows the training evolution through the epochs for the model $D=2$. The value of validation and training mean fidelity improves consistently across epochs, without indication of overfitting \cite{deeplearningbook}. The model maintains a relatively stable value after epoch $40$. This behavior is similar across all the trained models. 

Our experimental setup allows the acquisition of a large number of images for the training dataset. However, this fact does not always hold true for other experiments. Hence it is desirable to estimate the minimum number of experimental images, in order to train the network while the model still attains a prediction with acceptable mean fidelity. To do so, we train models for $D=2$, with training and validation datasets composed of $9000$ samples. For this approach, we generate theoretical samples through a computer-generated simulation. 

\begin{figure}[h!]
	\includegraphics[scale=0.6]{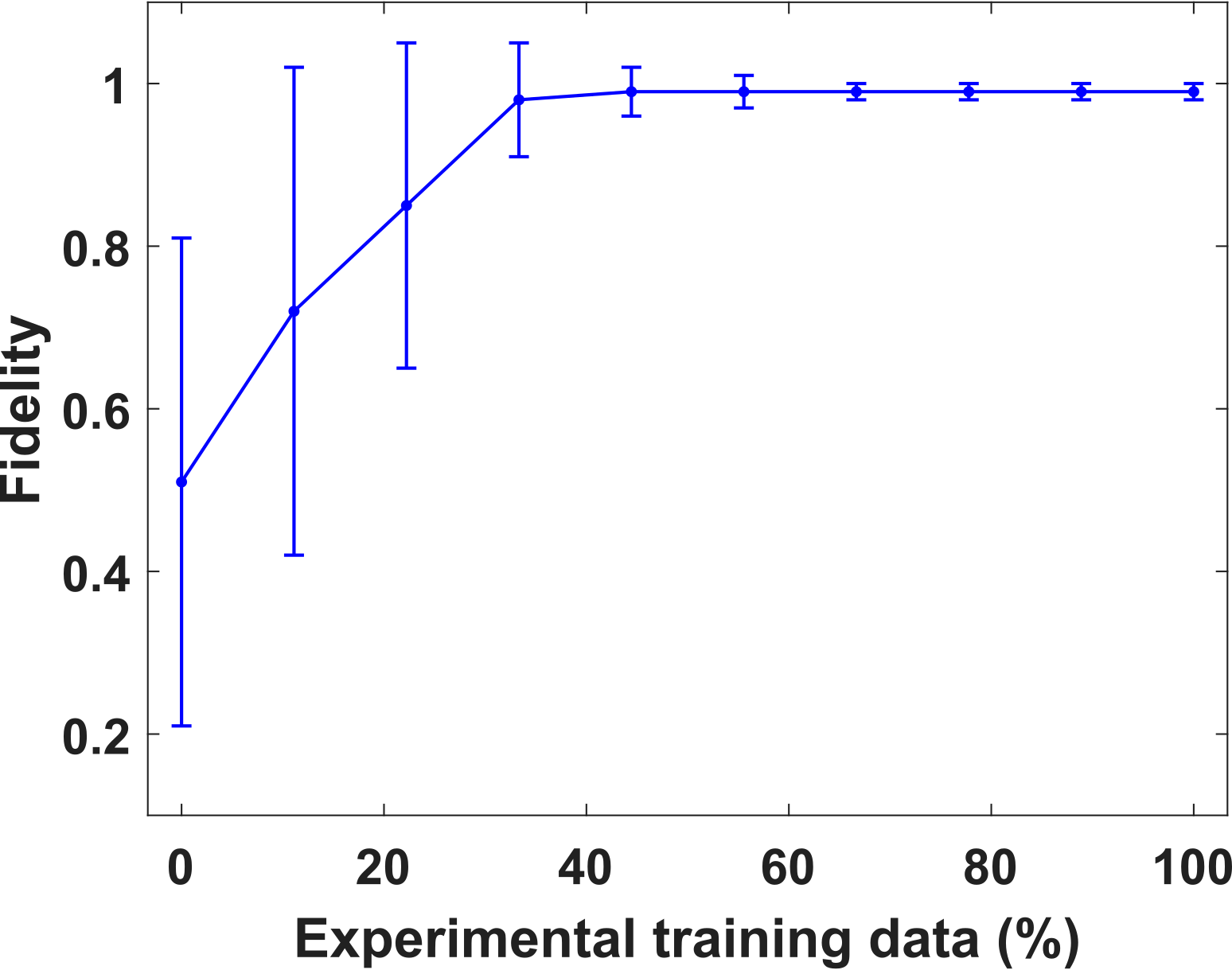}
	\caption{Mean fidelity as a function of the percentage of experimental train data.}
	\label{fig:d1s}
\end{figure}

Initially, we populate the entire dataset with theoretical samples. Then, we incrementally add experimental samples, maintaining a fixed dataset size. The mean fidelity of the superpositions inferred by each CNN models, as we increase the proportion of experimental samples, is shown in Fig. \ref{fig:d1s}. We tested all the models against the same testing dataset of $1000$ experimental samples.

If we use only theoretical images in the CNN's training and validation process, we obtain low mean fidelity  and a significant standard deviation ($0.5\pm 0.3$). Therefore, we observed that it is essential to use experimental images in the training of the network. We found that the minimum proportion of experimental data is $44\,\%$ for this specific problem. With this proportion, the model achieves a mean fidelity of $0.99 \pm 0.03$.

At a proportion of $66\,\%$ experimental images, the model achieve the fidelity of $0.99\pm0.01$. Above this percentage we have minimal gains in the performance of the model. We speculate that a the amount of data in our experimental dataset contains sufficient information for training the network even when only $66\,\%$ of the experimental samples are used. The result of mixing experimental and theoretical samples shows that it is possible to reach a reasonable estimation even for cases where it is not feasible to obtain a higher number of experimental samples. Since our experimental setup allows fast acquisition of samples, we choose to use exclusively experimental images in this work.

To verify our method, we train and test several CNN models, one for each dimension $D=2,3,4,5$, as shown in table \ref{table:1}. In this test, we inform the dimension of the mode superposition in order to select the correct CNN model. The mean fidelity concerning all superpositions in the testing dataset is calculated and the standard deviation gives the estimation error. We obtained a high mean fidelity value for all the dimensions analyzed. The error slightly increases for dimensions higher than $3$. The model is capable of performing the recognition in real-time, with an inference time of $0.9\, ms$ for a single superposition using a consumer-grade GPU (NVIDIA\textsuperscript{\textregistered} Geforce\textsuperscript{\textregistered} RTX 2080 Super\textsuperscript{TM}).

\begin{table}[h!]
\centering
 \begin{tabular}{||c| c| c||} 
 \hline
 Superposition  & Testing &  Mean  \\ 
 dimension & dataset & fidelity\\ [0.5ex] 
 \hline\hline
 2 & 1000 & 0.99 $\pm$ 0.01  \\ 
 \hline
 3 & 2000 & 0.99 $\pm$ 0.01  \\
 \hline
 4 & 3000 & 0.99 $\pm$ 0.02  \\
 \hline
 5 & 4000 & 0.99 $\pm$ 0.03  \\
 \hline
\end{tabular}
\caption{Testing dataset size and mean fidelity for each dimension.}
\label{table:1}
\end{table}

Quite remarkably, our system is also capable to determine the dimension of the mode superposition. The recognition of a superposition with an arbitrary, unknown order ($D \leq 5$) is demonstrated. First, the system tests the input experimental image against all the trained models. Then, the value of the estimated $c_{l,p}$ for each model is employed to generate the theoretical images. The superposition order is determined by comparing the theoretical images with the input image and selecting the model for which the inference produces the theoretical image that is most similar to the experimental input. We perform a blind test of the system by not informing the superposition order. The testing dataset contains $250$ superpositions of each order ($D =2, 3, 4, 5$), amounting to $1000$ samples. In this test, the recognition system gives the superposition coefficients and the dimension as outputs. The system achieved an accuracy of $99.7\,\%$ for the dimension estimation, and a mean fidelity of $0.99 \pm 0.02$.

In conclusion, we developed a tomographic method for the characterization of OAM superpositions based on two measurements and processed via machine learning. To define each superposition unequivocally, we perform two intensity measurements; the first is the direct image and the second is the image after applying an astigmatic transformation with a tilted lens. Once we have the two images, we use a convolutional neural network to recover the superposition coefficients. As we have shown, in cases where the experimental setup has limitations, it is possible to use theoretical images to increase the dataset. Nevertheless, to obtain a reasonable mean fidelity with a tolerable error, the minimum percentage of experimental images in the total training dataset is $44\,\%$. 
Our method was tested for $D=2,3,4,5$ using the experimental dataset. The results exhibit a high mean fidelity and low error, demonstrating that our model is reliable in different OAM space dimensions. In the last test, we did not inform the superposition dimension for the recognition system. Still, the method proved to be highly accurate, providing outstanding fidelity values and precise estimation of the dimension. Our method has a fast inference time, effectively enabling real-time recognition of superpositions with arbitrary order.

Finally, our method can be adapted to the quantum regime with a single-photon sensitive camera (such as an intensified or electron multiplying charge coupled device). Although image reconstruction requires a large number of photons to be gathered, the method can be supplemented by compression techniques to reconstruct images from a small number of detected photons per pixel. Moreover, the use of heralded single-photon sources can further improve the signal-to-noise ratio.

\section*{Acknowledgments}
Funding was provided by 
Coordena\c c\~{a}o de Aperfei\c coamento de Pessoal de N\'\i vel Superior (CAPES), 
Funda\c c\~{a}o Carlos Chagas Filho de Amparo \`{a} Pesquisa do Estado do Rio de Janeiro (FAPERJ), 
Funda\c c\~{a}o de Amparo \`{a} Pesquisa do Estado de Santa Catarina (FAPESC),
Conselho Nacional de Desenvolvimento Cient\'{\i}fico e Tecnol\'ogico (CNPq), 
and Instituto Nacional de Ci\^encia e Tecnologia de Informa\c c\~ao Qu\^antica 
(INCT-IQ 465469/2014-0).

\bibliography{bibliography}

\end{document}